\documentclass[oneside, draftcls, onecolumn, 12pt]{IEEEtran}

\usepackage{enumerate}
\usepackage{footnote} 
\usepackage{color}
\usepackage{url}
\usepackage{cite}
\usepackage{graphicx}
\usepackage{graphics}
\usepackage{psfrag}
\usepackage{array}
\usepackage{mdwtab}
\usepackage{epsfig}
\usepackage{subfigure}
\usepackage{amsmath,amssymb,amsfonts,mathrsfs,amscd}
\usepackage{pst-node,pst-text,pst-3d}
\usepackage{pstricks}

\newtheorem{remark}{Remark}
\newtheorem{proposition}{Proposition}
\newtheorem{definition}{Definition}
\newtheorem{theorem}{Theorem}

\newtheorem{lemma}{Lemma}

\newcommand{\sss}{\scriptscriptstyle}
\newcommand{\mbs}[1]{\pmb{#1}}
\newcommand{\vect}[1]{{\lowercase{\mbs{#1}}}}
\newcommand{\mat}[1]{{\uppercase{\mbs{#1}}}}

\newcommand{\bb}[1]{\mathbb{#1}}
\newcommand{\nnb}{\nonumber}

\newcommand{\Fig}[1]{Fig.~\!\ref{#1}}

\newcommand{\Eq}[1]{(\ref{#1})}

\newcommand{\ie}{\emph{i.e.}}

\newcommand{\D}{\displaystyle}

\renewcommand{\matrix}[1]{\begin{bmatrix}#1\end{bmatrix}}

\newcommand\Det{\mathrm{Det}}
\newcommand\transcsymbol{\scriptscriptstyle \dag \!}
\newcommand\transsymbol{\scriptscriptstyle \mathsf{T} \!}

\newcommand\Abs[1]{\left|#1\right|}

\newcommand\Abssqr[1]{\left|#1\right|^2}

\newcommand{\inv}[1]{{#1}^{\scriptscriptstyle -1 \!}}
\newcommand{\transc}[1]{{#1}^{\transcsymbol}}
\newcommand{\trans}[1]{{#1}^{\transsymbol}}
\newcommand{\pstv}[1]{{#1}^{\sss +}}

\newcommand\Norm[1]{\left\|{#1}\right\|}

\newcommand\defeq{\triangleq}

\newcommand{\Id}{\mathbf{I}}
\newcommand{\CN}[1][\Id]{\Ccal\Ncal\!\left(0,#1\right)}
\newcommand{\SNR}{{\mathsf{SNR}} }

\newcommand\iid{i.i.d.\ }

\newcommand\Pe{P_{\textrm{e}}}
\newcommand\Pout{P_{\textrm{out}}}

\newcommand{\asympteq}{\doteq} %

\newcommand{\nT}{n_\text{T}}
\newcommand{\nR}{n_\text{R}}
\newcommand{\nS}{n_\text{S}}
\newcommand{\PhiT}{\mPhi_\text{T}}
\newcommand{\PhiR}{\mPhi_\text{R}}
\newcommand{\PhiS}{\mPhi_\text{S}}

\newcommand\CC{\bb{C}}
\newcommand\EE{\bb{E}}

\newcommand\Ccal{\mathcal{C}}

\newcommand\Ncal{\mathcal{N}}
\newcommand\Ocal{\mathcal{O}}

\newcommand\Wcal{\mathcal{W}}

\newcommand{\mA}{\mat{A}}
\newcommand{\mB}{\mat{B}}

\newcommand{\mG}{\mat{G}}
\newcommand{\mH}{\mat{H}}

\newcommand{\mM}{\mat{M}}

\newcommand{\mP}{\mat{P}}
\newcommand{\mQ}{\mat{Q}}

\newcommand{\mT}{\mat{T}}
\newcommand{\mU}{\mat{U}}
\newcommand{\mV}{\mat{V}}
\newcommand{\mW}{\mat{W}}

\newcommand{\mv}{\vect{v}}

\newcommand{\mx}{\vect{x}}
\newcommand{\my}{\vect{y}}
\newcommand{\mz}{\vect{z}}

\newcommand{\mXi}{\mbs{\Xi}}

\newcommand{\mSigma}{\mbs{\Sigma}}

\newcommand{\mPhi}{\mbs{\Phi}}

\newcommand{\mlambda}{\boldsymbol{\lambda}}
\newcommand{\malpha}{\boldsymbol{\alpha}}

\newcommand{\mmu}{\boldsymbol{\mu}}
\newcommand{\mbeta}{\boldsymbol{\beta}}

\title{Diversity-Multiplexing Tradeoff of Double Scattering MIMO Channels \thanks{Manuscript submitted to the IEEE
     Transactions on Information Theory.  The authors
     are with the Department of Communications and Electronics,
     \'{E}cole Nationale Sup\'{e}rieure des T\'{e}l\'{e}communications, 46, rue Barrault, 75013
     Paris, France~(e-mail: syang@enst.fr; belfiore@enst.fr).}
}
 \author{%
 \authorblockN{Sheng Yang and Jean-Claude Belfiore}}

\begin{document}
\maketitle

\begin{abstract}
  It is well known that the presence of double scattering degrades the
  performance of a MIMO channel, in terms of both the multiplexing
  gain and the diversity gain. In this paper, a closed-form expression
  of the diversity-multiplexing tradeoff~(DMT) of double scattering
  MIMO channels is obtained. It is shown that, for a channel with
  $\nT$ transmit antennas, $\nR$ receive antennas and $\nS$
  scatterers, the DMT only depends on the ordered version of the
  triple $(\nT,\nS,\nR)$, for arbitrary $\nT, \nS$ and $\nR$.  The
  condition under which the double scattering channel has the same DMT as the single scattering channel is also established. 
\end{abstract}

\section{Introduction and Problem Description}
\label{sec:intro}
Multiple antennas are known as an important means to increase channel
capacity and to mitigate channel fadings~\cite{Foschini,Telatar}. The
tradeoff between the multiplexing gain and the diversity gain for
Rayleigh MIMO channels in the high SNR regime is characterized by the
diversity-multiplexing tradeoff~(DMT) proposed by Zheng and
Tse~\cite{Zheng_Tse}. However, the independent and identically
distributed~(i.i.d.) Gaussian property of the entries of MIMO channels is only
established under an idealistic assumption.  Recently, a more general model which
shows the scattering structure in the propagation environment has been
proposed~\cite{Gesbert}.  This model considers the rank deficiency as
well as the fading correlation, by characterizing the channel matrix
as a product of two statistically independent complex Gaussian
matrices.

The presence of double scattering degrades considerably the
performance promised by MIMO channels, for both the multiplexing gain
and the diversity gain. Intuitively, the performance of double
scattering MIMO channels is not better than either the
transmitter-scatterers or the scatterers-receiver channel. One
interesting question is: \emph{``what is the impact of double
scattering on the channel's capability of capturing diversity and
providing multiplexing gain in the high SNR regime~?''}. This question
is answered in this work, by studying the DMT of double scattering
MIMO channels.

More precisely, the main contribution of this work is to provide a
closed-form expression of the DMT for general double-scattering MIMO
channels. It is shown that, for a MIMO channel with $\nT$ transmit
antennas, $\nS$ scatterers and $\nR$ receive antennas, the DMT only
depends on the ordered triple of $(\nT, \nS, \nR)$. This property can
be seen as a generalization of the reciprocity of MIMO channels. It is
also shown that the upperbound on the channel diversity order $\nT \nS
\nR/\max\left\{\nT,\nS,\nR\right\}$ is usually not achievable, unless
for $(\nT,\nS,\nR)$ satisfying 
\begin{equation*}
        2\max\left\{\nT,\nS,\nR\right\}+1\geq\nT+\nS+\nR.
\end{equation*}%

In this paper, we use boldface lower case letters $\mbs{v}$ to denote
vectors, boldface capital letters $\mbs{M}$ to denote matrices.
$\Ccal\Ncal$ represents the complex Gaussian random variable.
$\trans{[\cdot]},\transc{[\cdot]}$ respectively denote the matrix
transposition and conjugated transposition operations. $\Norm{\cdot}$
is the vector norm. $\pstv{(x)}$ means $\max(0,x)$. $\Det(\mM)$ is the
absolute value of the determinant $\det(\mM)$. The square root
$\mP^{1/2}$ of a positive semi-definite matrix $\mP$ is defined as a
positive semi-definite matrix such that
$\mP=\mP^{1/2}\transc{\bigl(\mP^{1/2}\bigr)}$. The dot equal operator
$\asympteq$ denotes asymptotic equality in the high SNR regime, \ie,
\begin{equation*}
  p_1 \asympteq p_2 \quad \textrm{means} \quad
  \lim_{{\sss\textsf{SNR}}\to\infty}\frac{\log p_1}{\log \SNR} =
  \lim_{{\sss\textsf{SNR}}\to\infty}\frac{\log p_2}{\log \SNR}.
\end{equation*}%

The rest of the paper is organized as follows.
Section~\ref{sec:system-model} introduces the channel model, some
preliminaries on complex Wishart matrices and the
DMT. Section~\ref{sec:dmt-prod-Rayleigh} studies the DMT of Rayleigh
product channels, a particular case of the double scattering
channel. The DMT of a general double scattering channel is provided in
Section~\ref{sec:dmt-double-scatter}.  Section~\ref{sec:conclusion}
draws a brief conclusion on this work and the Appendix is dedicated to
some lemmas and their proofs.

\section{System Model and Preliminaries}
\label{sec:system-model}

\subsection{Channel Model}
\label{sec:channel-model}

In this paper, we consider the double scattering MIMO channel with
$\nT$ transmit antennas, $\nS$ scatterers and $\nR$ receive antennas
in the following form
\begin{equation}
  \label{eq:channel-model}
  \my = \sqrt{C\,{\SNR}} \mH \mx + \mz
\end{equation}%
with 
\begin{equation}
  \label{eq:channel-matrix}
  \mH \defeq \PhiR^{1/2}\mH_1\PhiS^{1/2}\mH_2\PhiT^{1/2}
\end{equation}%
where $\mx\in\CC^{\nT}$ is the transmitted signal with \iid unit
variance entries; $\my\in\CC^{\nR}$ represents the received signal;
$\mz\in\CC^{\nR}$ is the AWGN with $\mz\sim\CN[\Id]$; the constrant
$C$ is the normalization factor such that $\SNR$ is the average Signal
to Noise Ratio per receive antenna. $\mH_1 \in \CC^{\nR\times\nS}$ and
$\mH_2 \in \CC^{\nS\times\nT}$ are statistically independent matrices
with \iid unit variance Gaussian entries. Correlations at each node
are characterized by $\PhiT$, $\PhiS$ and $\PhiR$ which are assumed to
be positive definite matrices\footnote{The correlation matrices are
  positive semi-definite in general. However, it is always possible to
  have an equivalent channel model of positive definite $\mPhi$'s and
  Gaussian matrices $\mH_i$'s of reduced dimensions, using the
  eigenvalue decomposition of the correlation matrices and the
  unitarily invariance property of Gaussian matrices.  In this case,
  the effective numbers of antennas and scatterers are $\nT'$, $\nR'$
  and $\nS'$, \ie, the respective ranks of $\PhiT$, $\PhiR$ and
  $\PhiS$.} with respective dimensions $\nT\times \nT$, $\nS\times
\nS$ and $\nR\times \nR$. We denote such a channel, a $(\nT, \nS,
\nR)$ channel hereafter.

\subsection{Wishart Matrices}

\begin{definition}[Wishart Matrix]
  The $m\times m$ random matrix $\mW = \mH\transc{\mH}$ is a (central)
  complex Wishart matrix with $n$ degrees of freedom and covariance
  matrix $\mSigma$,~(denoted as $\mW\sim\Wcal_m(n,\mSigma)$), if the
  columns of the $m\times n$ matrix $\mH$ are zero-mean independent
  complex Gaussian vectors with covariance matrix $\mSigma$.
\end{definition}

\begin{theorem}[\!\!\cite{James,Gao_Smith,Simon,Tulino_Verdu}]\label{thm:Wishart}
  Let $\mW$ be a central complex Wishart matrix
  $\mW\sim\Wcal_m(n,\mSigma)$, where the eigenvalues of $\mSigma$ are
  distinct and their ordered values are $\mu_1>\ldots>\mu_m>0$. Let
  $\lambda_1>\ldots>\lambda_q>0$ be the ordered positive eigenvalues
  of $\mW$ with $q\defeq\min\{m,n\}$. The joint p.d.f. of $\mlambda$
  is
  \begin{equation}
    \label{eq:Wishart:n>m}
    K_{m,n} {\Det\left[e^{-\lambda_j/\mu_i}\right]} \prod_{i=1}^m \mu_i^{m-n-1} \lambda_i^{n-m} \prod_{i<j}^m \frac{\lambda_i-\lambda_j}{\mu_i-\mu_j} 
  \end{equation}%
  for $n\geq m$, and
  \begin{equation}
    \label{eq:Wishart:n<m}
    G_{m,n} {\Det(\mXi)} \prod_{i<j}^m \frac{1}{(\mu_i-\mu_j)} \prod_{i<j}^n (\lambda_i-\lambda_j) 
  \end{equation}%
  for $n<m$ with 
  \begin{equation}
    \label{eq:def-Xi}
    \mXi \defeq \matrix{1 & \mu_1 & \cdots & \mu_1^{m-n-1} & \mu_1^{m-n-1}e^{-\frac{\lambda_1}{\mu_1}} & \cdots & \mu_1^{m-n-1}e^{-\frac{\lambda_n}{\mu_1}} \\
      \vdots & \vdots & \ddots &\vdots& \vdots& \ddots & \vdots \\
        1 & \mu_m & \cdots & \mu_m^{m-n-1} & \mu_m^{m-n-1}e^{-\frac{\lambda_1}{\mu_m}} & \cdots & \mu_m^{m-n-1}e^{-\frac{\lambda_n}{\mu_m}}
}.
  \end{equation}%
  $K_{m,n}$ and $G_{m,n}$ are normalization factors. In particular,
  for $\mSigma=\Id$, the joint p.d.f. is
  \begin{equation}
    \label{eq:Wishart:Id}
    P_{m,n}  e^{-\sum_i \lambda_i} \prod_{i=1}^q \lambda_i^{\Abs{m-n}} \prod_{i<j}^q (\lambda_i-\lambda_j)^2.    
  \end{equation}
\end{theorem}

\subsection{Diversity-Multiplexing Tradeoff}
\label{sec:divers-mult-trad}

\begin{definition}[Multiplexing and diversity gains\cite{Zheng_Tse}]
  A coding scheme $\{\Ccal(\SNR)\}$ is said to achieve
  \emph{multiplexing gain} $r$ and \emph{diversity gain} $d$ if
\begin{equation*}
  \lim_{{\sss\textsf{SNR}}\to\infty} \frac{R(\SNR)}{\log\SNR} = r \quad
\textrm{and}\quad 
  \lim_{{\sss\textsf{SNR}}\to\infty} \frac{\log\Pe(\SNR)}{\log\SNR} = -d
\end{equation*}
where $R(\SNR)$ is the data rate measured by bits per channel use~(PCU)
and $\Pe(\SNR)$ is the average error probability using a maximum
likelihood~(ML) decoder.
\end{definition}
For any linear fading Gaussian channel~
\begin{equation*}
  \my = \sqrt{\SNR}\,\mH\,\mx + \mz
\end{equation*}%
where $\mz$ is an AWGN with $\EE\bigl\{\mz\transc{\mz}\bigr\}=\Id$ and
$\mx$ is subject to the input power constraint
$\text{Tr}\left\{\EE\left[\mx\transc{\mx}\right]\right\}\leq 1$, the
DMT $d(r)$ can be found as the exponent of the outage probability
in the high SNR regime, \ie,
\begin{align}
  \Pout(r\log\SNR) &\asympteq \textrm{Prob}\bigl\{\log\det\left(\Id+\SNR\,\mH\transc{\mH}\right)\leq r\log\SNR \bigr\} \nnb\\
  &= \textrm{Prob}\bigl\{\det\left(\Id+\SNR\,\mH\transc{\mH}\right)\leq \SNR^r \bigr\} \nnb\\
  &\asympteq \SNR^{-d(r)}. \label{eq:dmt}
\end{align}

\begin{lemma}[Calculation of diversity-multiplexing tradeoff]
  \label{lemma:cal-dmt}
  Consider a linear fading Gaussian channel defined by $\mH$ for which
  $\det\left(\Id+\SNR\,\mH\transc{\mH}\right)$ is a function of $\mv$,
  a vector of positive random variables. Then, the DMT $d(r)$ of this
  channel can be calculated as
  \begin{equation*}
    d(r) = \inf_{\Ocal(\malpha, r)} \varepsilon(\malpha)
  \end{equation*}%
  where $\alpha_i\defeq-\log v_i/\log\SNR$ is the exponent of $v_i$,
  $\Ocal(\malpha, r)$ is the outage event set in terms of $\malpha$
  and $r$ in the high SNR regime, and $\varepsilon(\malpha)$ is the
  exponential order of the p.d.f. $p_{\malpha}(\malpha)$ of $\malpha$,
  \ie,
  \begin{equation*}
    p_{\malpha}(\malpha) \asympteq \SNR^{-\varepsilon(\malpha)}.
  \end{equation*}%
\end{lemma}%
\begin{proof}
  This lemma is justified by \Eq{eq:dmt} using Laplace's method,
  as shown in \cite{Zheng_Tse}.
\end{proof}

As an example, the DMT of an $\nR \times \nT$ Rayleigh MIMO channel is
a piecewise-linear function connecting the points
$(k,d(k)),k=0,1,\ldots,\min\{\nR,\nT\}$, where~\cite{Zheng_Tse}
\begin{equation}
  \label{eq:DMT-Rayleigh}
  d(k) = (\nR-k)(\nT-k).
\end{equation}%

\section{Diversity-Multiplexing Tradeoff of Rayleigh Product Channels}  
\label{sec:dmt-prod-Rayleigh}
In this section, we study a special case of the double scattering MIMO
channel, where $\PhiT, \PhiS$ and $\PhiR$ are identity matrices. We
call it a Rayleigh product channel.

\begin{theorem}\label{thm:DMT-prod-Rayleigh}
  Let $\mH\defeq\mH_2\mH_1$ with $\mH_2 \in \CC^{n\times l}$ and 
$\mH_1 \in \CC^{l\times m}$ being independent Gaussian matrices with
  \iid $\CN[1]$ entries. Define $(M,N,L)$ be the ordered version of
  $(m,n,l)$ with $M\leq N\leq L$. Then, the diversity-multiplexing
  tradeoff of the fading channel 
  \begin{equation*}
    \my = \sqrt{\frac{\SNR}{l\,m}} \mH \mx + \mz 
  \end{equation*}%
  is a piecewise-linear function connecting the points
  $(k,d(k)),k=0,\ldots,M$, where
  \begin{equation}
    \label{eq:DMT-Prod-Rayleigh}
    d(k) = (M-k)(N-k) - \left\lfloor \frac{\left[(M-\Delta-k)^+\right]^2}{4}\right\rfloor    
  \end{equation}%
  with $\Delta\defeq L-N$.  
\end{theorem}
Before going to the proof, some remarks can be made about the DMT of a
Rayleigh product channel.
\begin{remark}\label{remark:1}
  From \Eq{eq:DMT-Prod-Rayleigh}, we note that
  \begin{enumerate}
  \item The DMT does not depend on the triple $(m,n,l)$ but only on
    the ordered triple $(M,N,L)$, which can be seen as a
    generalization of the reciprocity property~\cite{Telatar} of MIMO
    channels;
  \item The DMT of a Rayleigh product channel is always inferior to
    that of an $M\times N$ Rayleigh channel, \ie, $d(k)$ is
    upperbounded by $\bar{d}(k)\defeq(M-k)(N-k)$;
  \item The upperbound $\bar{d}(k)$ is achieved for $k\geq
    M-\Delta-1$, which means that $d(k)$ coincides with $\bar{d}(k)$
    at least for the last section of the curve;
  \item When $L+1\geq M+N$, the Rayleigh product channel has exactly
    the same DMT performance as an $M\times N$ Rayleigh channel;
  \item Finally, as a consequence of the previous observation, a
    Rayleigh product channel is always equivalent to an $N\times 1$
    Rayleigh channel when $M=1$.
  \end{enumerate}
\end{remark}
We should point out that the relation between the Gaussian coding
bound and the outage bound studied in \cite{Zheng_Tse} is intimately
related to the Rayleigh product channel. In \cite{Zheng_Tse}, it is
shown that the Gaussian codeword matrix should be long enough to
achieve the DMT of the Rayleigh MIMO channel. The code length
condition is exactly the same as the condition provided by
observation~4 in the remark above.

As in \cite{Zheng_Tse}, the DMT is obtained from the p.d.f.  of the
eigenvalues of $\mQ_{\mH}\defeq\mH\transc{\mH}$, which depends on
$(m,n,l)$. For now, we know that $\mQ_1\defeq \mH_1\transc{\mH}_1 \sim
\Wcal_l(m,\Id)$. Let us define the eigenvalues of $\mQ_1$ as
$\mu_1>\ldots>\mu_{\min\{l,m\}}$. Then,
$\mQ_{\mH}=\left(\mH_2\mQ_1^{1/2}\right)\transc{\left(\mH_2\mQ_1^{1/2}\right)}$
has the same eigenvalues as $\mQ_{\mG}\defeq\mG\transc{\mG}$ with
$\mG\defeq \mQ_1^{1/2} \transc{\mH}_2$. By definition, conditionned on
$\mH_1$, we have $\mQ_{\mG}\sim\Wcal_l(n,\mQ_1)$. Therefore, from now
on, we can study the eigenvalues
$\lambda_1>\ldots>\lambda_{\min\{l,m,n\}}$of $\mQ_{\mG}$, whose joint
p.d.f.  only depends on the eigenvalues of $\mQ_1$, according to
Theorem~\ref{thm:Wishart}. In the rest of this section, we prove
Theorem~\ref{thm:DMT-prod-Rayleigh} in two cases~: $\min\{m,n\}\geq l$
and $\min\{m,n\} < l$.

\subsection{The $\min\{m,n\} \geq l$ Case}
\label{sec:minm-n-geq}
In this case, we can exchange $m$ and $n$, by the reciprocity property
of MIMO channels. Without loss of generality, we assume that $m\geq
n$.  From \Eq{eq:Wishart:n>m} and \Eq{eq:Wishart:Id}, we get the joint
p.d.f. of $(\mlambda,\mmu)$
\begin{equation*}
  \begin{split}
  p_{\mlambda,\mmu}(\mlambda,\mmu) &= C_{l,m,n} \prod_{i=1}^l \mu_i^{m-n-1}\lambda_i^{n-l} \prod_{i<j}^l {(\lambda_i-\lambda_j)}{(\mu_i-\mu_j)} \\
  &\quad \cdot \exp\left(-\sum_{i=1}^l \mu_i\right)
  \Det\left[e^{-\lambda_j/\mu_i}\right],    
  \end{split}
\end{equation*}
where $C_{l,m,n}$ is the normalization factor. Define
$\alpha_i\defeq-\log\lambda_i/\log\SNR $ and
$\beta_i\defeq-\log\mu_i/\log\SNR$ for $i=1,\ldots,l$. Then, we have
\begin{equation}
  \label{eq:joint-pdf-ab}
  \begin{split}
    p_{\malpha,\mbeta}(\malpha,\mbeta) &= C_{l,m,n} (\log\SNR)^{2l}\prod_{i=1}^l \SNR^{-(n-l+1)\alpha_i} \SNR^{-(m-n)\beta_i} \nnb\\
    &\quad\cdot \prod_{i<j}^l {(\SNR^{-\alpha_i}-\SNR^{-\alpha_j})}{(\SNR^{-\beta_i}-\SNR^{-\beta_j})}  \nnb\\
    &\quad \cdot \exp\left(-\sum_{i=1}^l \SNR^{-\beta_i}\right)
    \Det\left[\exp\left(-\SNR^{-(\alpha_j-\beta_i)}\right)\right].
  \end{split}
\end{equation}
First, we only consider $\beta_i\geq0,\forall i$, since otherwise,
$\exp\left(-\sum_{i} \SNR^{-\beta_i}\right)$ would decay exponentially
with $\SNR$\cite{Zheng_Tse}. The high SNR exponent of the quantity
$\Det\left[\exp\left(-\SNR^{-(\alpha_j-\beta_i)}\right)\right]$ is
calculated in Lemma~\ref{lemma:lemma1}. From \Eq{eq:detexp}, we only
need to consider $\alpha_i\geq\beta_i,\forall i$, so that
$p_{\malpha,\mbeta}(\malpha,\mbeta)$ does not decay exponentially.
Finally, by Lemma~\ref{lemma:cal-dmt}, the DMT $d(r)$ can be obtained
by solving the optimization problem
\begin{equation}\label{eq:optpbl}
  d(r) = \inf_{\Ocal(\malpha,\mbeta,r)} \epsilon(\malpha,\mbeta)
\end{equation}%
with
\begin{equation*}
   \Ocal(\malpha,\mbeta,r) \defeq \left\{(\malpha,\mbeta):\quad\sum_{i=1}^l(1-\alpha_i)^+ < r, {{\alpha_1\leq\cdots\leq\alpha_l,\atop\beta_1\leq\cdots\leq\beta_l},\alpha_i\geq\beta_i\geq0,\forall i}\right\}
\end{equation*}
and 
\begin{equation}\label{eq:epsilon:ab}
  \epsilon(\malpha,\mbeta) \defeq \sum_{i=1}^l (n-i+1)\alpha_i + \sum_{i=1}^l (m-n+l-i)\beta_i + \sum_{i<j}^l (\alpha_i-\beta_j)^+. 
\end{equation}%

\begin{figure*}
  \subfigure[The $\min\{m,n\}\geq l$ case]{
  \epsfig{figure=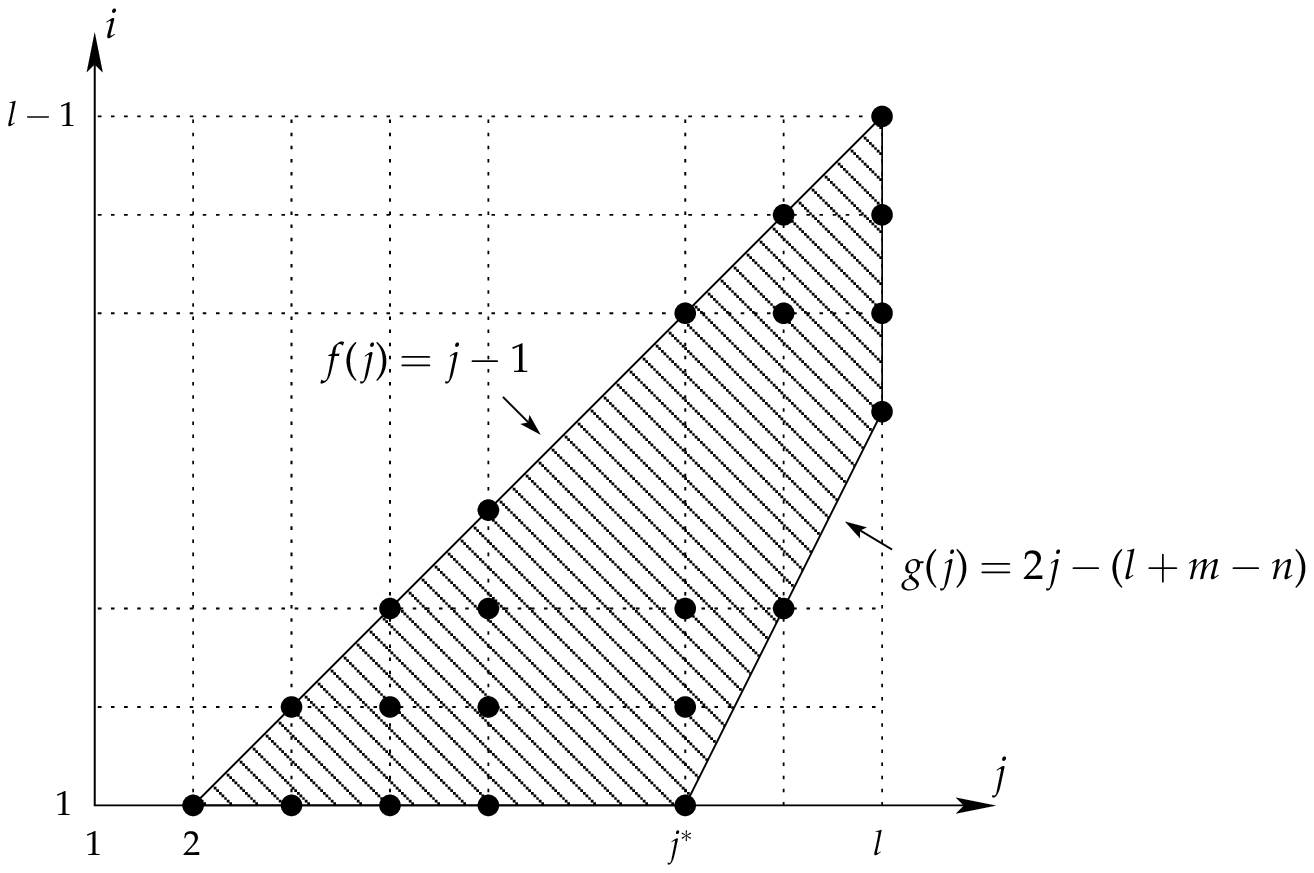,width=0.525\textwidth}
  \label{fig:find-ci}    }
  \subfigure[The $\min\{m,n\} < l$ case]{
  \epsfig{figure=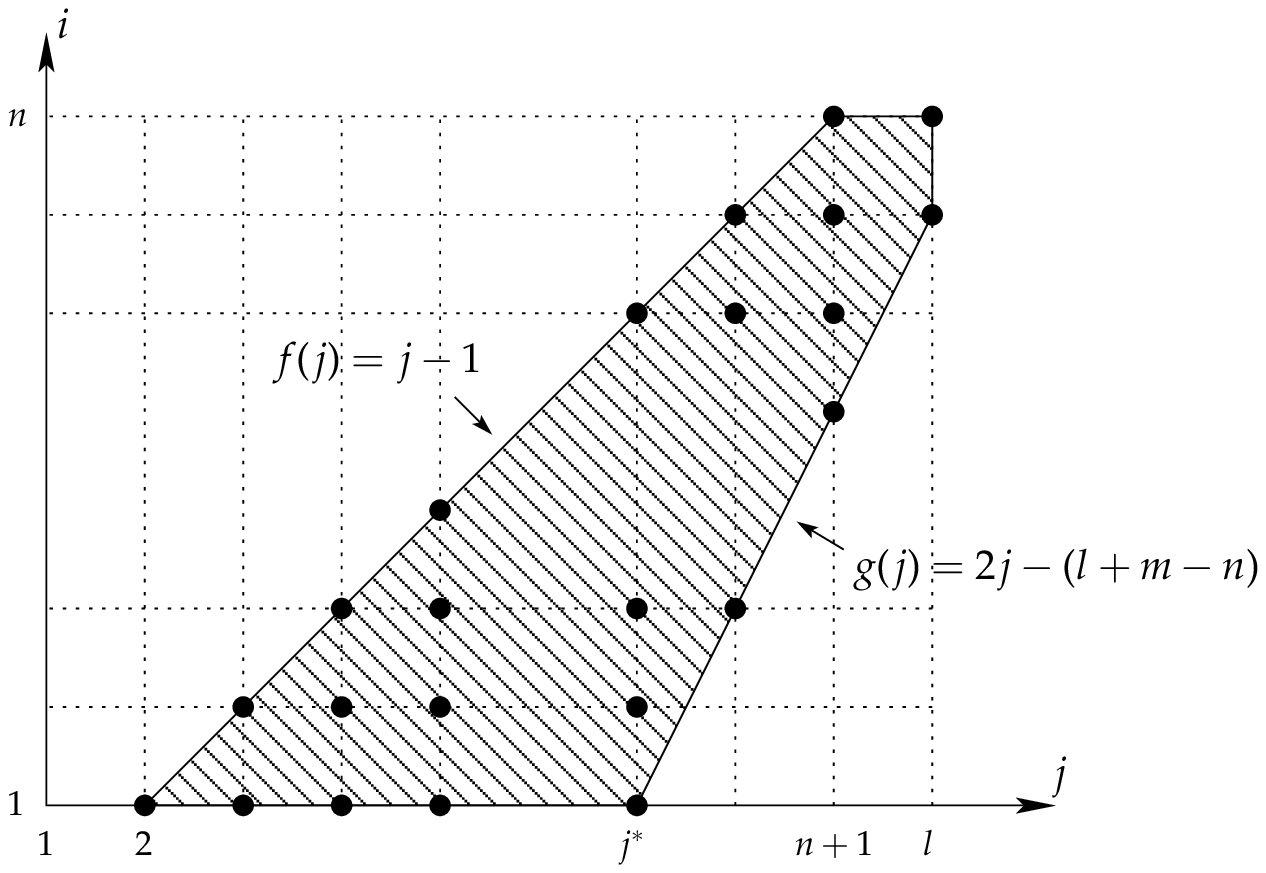,width=0.5\textwidth}
  \label{fig:find-ci2}  }
\caption{For each $j$, the black dots represent the $\alpha$'s that are freed by $\beta_j$. For each $i$, 
  the number of black dots $c_i$ is the coefficient of $\alpha_i$.
  Thus, for $i\leq g(l)$, $c_i=\left\lfloor{g^{-1}(i)}\right\rfloor -
  \left\lceil{ f^{-1}(i) }\right\rceil + 1$; and for $i>g(l)$, $c_i=l
  - \left\lceil{ f^{-1}(i) }\right\rceil + 1$.}
\end{figure*}

The optimization problem \Eq{eq:optpbl} can be solved in two steps: 1)
find optimal $\mbeta$ by fixing $\malpha$, and then 2) optimize
$\malpha$. Let us start from the feasible region
\begin{equation} \label{eq:tmp1}
  0 \leq \beta_1 = \alpha_1 \leq \beta_2 = \alpha_2 \leq \cdots \leq \beta_l= \alpha_l
\end{equation}%
in which we have $\D\sum_{i<j}^l(\alpha_i-\beta_j)^+ = 0$. Note that
for each $j$, the feasibility conditions require that $\beta_j$ should
only move to the left in terms of its positions\footnote{The position
  here refers to the position in the inequality chain of $\alpha_i$'s
  and $\beta_i$'s in increasing order, as the one in \Eq{eq:tmp1}.}
relative to the $\alpha_i$'s and that $\beta_i$ should never be on the
left of $\beta_j$ for $i>j$. Each time $\beta_j$ passes an $\alpha_i$
from right to left, $\sum_{i<j}(\alpha_i-\beta_j)^+$ increases by
$\alpha_i-\beta_j$, which increases the coefficient of $\alpha_i$ by
$1$ and decreases the coefficient of $\beta_j$ by $1$. To minimize the
value of $\epsilon(\malpha,\mbeta)$, $\beta_j$ is allowed to pass
$\alpha_i$ only when the current coefficient of $\beta_j$ in
\Eq{eq:epsilon:ab} is positive\footnote{When the coefficient of
  $\beta_j$ in \Eq{eq:epsilon:ab} is positive, decreasing $\beta_j$
  decreases $\epsilon(\malpha,\mbeta)$.}. The maximum number of
$\alpha_i$ that can be ``freed'' by $\beta_j$ is $j-1$, \ie,
$\alpha_{j-1},\ldots,\alpha_1$. Note that the initial coefficient of
$\beta_j$ is $m-n+l-j$ and is decreasing with $j$ while the number
$j-1$ is increasing with $j$. Let $j^*$ be the largest number such
that $m-n+l-j\geq j-1$. Obviously, for $j\leq j^*$,
$\alpha_{j-1},\ldots,\alpha_1$ can be freed and the final coefficients
of $\beta_j$ is $m-n+l-2j-1$~($\geq 0$) and $\beta_j^*=0$. For
$j>j^*$, $\beta_j$ can only free
$\alpha_{j-1},\ldots,\alpha_{j-(m-n+l-j)}$ and the final coefficient
of $\beta_j$ is $0$. Substituting the optimal solutions $\beta_j^*$'s
back into \Eq{eq:epsilon:ab}, we get
\begin{equation}
  \label{eq:epsilon:a}
  \epsilon(\malpha) = \sum_{i=1}^l (n-i+1+c_i) \alpha_i
\end{equation}%
where $c_i$ can be found with the help of \Fig{fig:find-ci}. Finally,
we have
\begin{equation*}
  \begin{split}
    \epsilon(\malpha) &= \sum_{i=1}^{l-(m-n)} \left(n+1-2i+\left\lfloor\frac{l+i+(m-n)}{2}\right\rfloor\right)\alpha_i \\
    &\quad + \sum_{i=l-(m-n)+1}^{l} \left(n+l+1-2i\right) \alpha_i 
  \end{split}
\end{equation*}
where the coefficient of $\alpha_i$ is non-negative and is
non-increasing with $i$. Hence, the optimal solution is
$\alpha_i^*=1,i=k+1,\ldots,l$ and $\alpha_i^*=0,i=1,\ldots,k$, from
which we can verify that
\begin{equation}
  \label{eq:DMT-Prod-Rayleigh1}
  d(k) = (l-k)(n-k) - \left\lfloor \frac{\left[(l-(m-n)-k)^+\right]^2}{4}\right\rfloor.    
\end{equation}%

\subsection{The $\min\{m,n\} <  l$ Case}
\label{sec:minm-n-le}
Again, by the reciprocity property, we assume that $n\leq m$. However,
we should study the $m\geq l$ case and the $m<l$ case separately. We
start with the former case.

\subsubsection{The $n<l\leq m$ Case}
\label{sec:nlleq-m-case}

From \Eq{eq:Wishart:n<m} and \Eq{eq:Wishart:Id}, we get the joint
p.d.f. of $(\mlambda,\mmu)$
\begin{equation}
  \label{eq:joint-pdf2}
  \begin{split}
    p_{\mlambda,\mmu}(\mlambda,\mmu) &= B_{l,m,n} \prod_{i=1}^l
    \mu_i^{m-l} \prod_{i<j}^l{(\mu_i-\mu_j)} \\ 
    &\quad\cdot\prod_{i<j}^n {(\lambda_i-\lambda_j)} \,\Det\left(\mXi\right)    
  \end{split}
\end{equation}
where $B_{l,m,n}$ is the normalization factor. Same procedure as the
previous case and Lemma~\ref{lemma:lemma2} lead to the following
asymptotical p.d.f. of $(\malpha,\mbeta)$
\begin{equation*}
  \begin{split}
  p_{\malpha,\mbeta}(\malpha,\mbeta) &\asympteq (\log\SNR)^{l+n}\prod_{i=1}^n \SNR^{-(n-i+1)\alpha_i} \\
  &\quad\cdot \prod_{i=1}^{n+1}\SNR^{-(l+m-n-i)\alpha_i} \prod_{i=n+2}^{l}\SNR^{-(l+m+1-2i)\alpha_i} \\
  &\quad \cdot \prod_{i=1}^{n}\prod_{j=n+1}^{l}\SNR^{-(\alpha_i-\beta_j)^+}  \prod_{i<j}^n \SNR^{-(\alpha_i-\beta_j)^+} \\
  &\quad\cdot \exp\left(-\sum_{i=1}^l \SNR^{-\beta_i}\right)
  \exp\left(-\sum_{i=1}^n \SNR^{-(\alpha_i-\beta_i)}\right).
  \end{split}
\end{equation*}
As before, we only consider $\beta_i\geq0,\forall i$, and
$\alpha_i\geq\beta_i$, for $i=1,\ldots,n$, in order that
$p_{\malpha,\mbeta}(\malpha,\mbeta)$ does not decay exponentially.
Finally, the DMT $d(r)$ can be obtained by solving the optimization
problem \Eq{eq:optpbl} with
\begin{equation*}
   \Ocal(\malpha,\mbeta,r) \defeq \left\{(\malpha,\mbeta):\quad\sum_{i=1}^n(1-\alpha_i)^+ < r, {{\alpha_1\leq\cdots\leq\alpha_n,\atop\beta_1\leq\cdots\leq\beta_l},\alpha_i\geq\beta_i\geq0,\text{for}\ i=1,\ldots,n}\right\}
\end{equation*}
and 
\begin{equation}
  \label{eq:epsilon:ab2}  
  \begin{split}
    \epsilon(\malpha,\mbeta) =& \sum_{i=1}^n (n+1-i)\alpha_i + \sum_{i=1}^{n+1} (l+m-n-i)\beta_i\\
    &\quad + \sum_{i=n+2}^l (l+m+1-2i)\beta_i + \sum_{i=1}^n\!\sum_{j=n+1}^l (\alpha_i-\beta_j)^+ \\
    &\quad + \sum_{i<j}^n (\alpha_i-\beta_j)^+.
  \end{split}
\end{equation}
The optimization procedure is exactly the same as in the previous case.
With the optimal $\beta_j$'s, we have
\begin{equation}
  \label{eq:epsilon:a2}
  \epsilon(\malpha) = \sum_{i=1}^n (n-i+1+c_i) \alpha_i
\end{equation}%
where $c_i$ can be found with the help of \Fig{fig:find-ci2}. Finally,
we have
\begin{equation*}
  \begin{split}
    \epsilon(\malpha) &= \sum_{i=1}^{l-(m-n)} \left(n+1-2i+\left\lfloor\frac{l+i+(m-n)}{2}\right\rfloor\right)\alpha_i \\ 
    &\quad + \sum_{i=l-(m-n)+1}^{n} \left(n+l+1-2i\right) \alpha_i 
  \end{split}
\end{equation*}
where the coefficient of $\alpha_i$ is non-negative and is
non-increasing with $i$. Hence, the optimal solution is
$\alpha_i^*=1,i=k+1,\ldots,l$ and $\alpha_i^*=0,i=1,\ldots,k$, from
which we have
\begin{equation}
  \label{eq:DMT-Prod-Rayleigh2}
  d(k) = (l-k)(n-k) - \left\lfloor \frac{\left[(n-(m-l)-k)^+\right]^2}{4}\right\rfloor.    
\end{equation}%

\subsubsection{The $n\leq m<l$ Case}
\label{sec:nlleq-m-case}

In this case, $\mu_{m+1}=\cdots=\mu_{l}=0$ with probability $1$. Let
$\mmu\defeq\trans{\left[\mu_1\cdots\mu_m\right]}$ be the vector of the
nonzero eigenvalues of $\mQ_1$. The conditional p.d.f.
$p_{\mlambda|\mmu}(\mlambda|\mmu)$ is given by
Lemma~\ref{lemma:lemma3}. The p.d.f. of $\mmu$ being known from
\Eq{eq:Wishart:Id}, we get the joint p.d.f. of $(\mlambda,\mmu)$ in
exactly the same form as \Eq{eq:joint-pdf2}, except that $l$ and $m$
are interchanged. We have directly
\begin{equation}
  \label{eq:DMT-Prod-Rayleigh3}
  d(k) = (m-k)(n-k) - \left\lfloor \frac{\left[(n-(l-m)-k)^+\right]^2}{4}\right\rfloor.
\end{equation}%

\section{Diversity-Multiplexing Tradeoff of Double Scattering MIMO Channels}  
\label{sec:dmt-double-scatter}
In this section, we study the DMT of a general double
scattering channel, where the antenna and scatterer correlations
$\PhiT$, $\PhiS$ and $\PhiR$ are non-trivial. 

It is intuitive to expect that the DMT is independent of the
correlation matrices, as long as they are not singular, since the DMT
is an asymptotical performance measure. First of all, it is easy to
show that the antenna correlations $\PhiT$ and $\PhiR$ do not
affect the tradeoff. To see this, note that
\begin{align}
  \det(\Id + \SNR \mH\transc{\mH})
  &= \det(\Id + \SNR \PhiR^{1/2} \mH_2 \PhiS^{1/2} \mH_1 \PhiT \transc{\mH}_1 \PhiS^{1/2}\transc{\mH}_2 \PhiR^{1/2}) \nnb\\
  &\asympteq \det(\Id + \SNR \PhiR^{1/2} \mH_2 \PhiS^{1/2} \mH_1 \transc{\mH}_1 \PhiS^{1/2}\transc{\mH}_2 \PhiR^{1/2}) \nnb\\
  &\asympteq \det(\Id + \SNR \transc{\mH}_1 \PhiS^{1/2}\transc{\mH}_2 \mH_2 \PhiS^{1/2} \mH_1 ) \nnb
\end{align}
where $\PhiT$ and $\PhiR$ disappear in the high SNR analysis. Now,
it remains to show that $\PhiS$ has no impact on the high SNR
analysis. The following proposition confirms this statement.
\begin{proposition}\label{prop:invariance-asymp}
  Let $\mM$ be any $m\times n$ random matrix and $\mT$ be any $m\times
  m$ non-singular matrix whose singular values satisfy 
  $\sigma_{\min}(\mT)\asympteq\sigma_{\max}(\mT)\asympteq\SNR^0$.
  Define $q\defeq\min\{m,n\}$ and $\mbs{\tilde{M}} \defeq \mT\mM$. Let
  $\sigma_1(\mM)>\ldots>\sigma_q(\mM)>0$ and
  $\sigma_1(\mbs{\tilde{M}})>\ldots>\sigma_q(\mbs{\tilde{M}})>0$ be
  the distinct ordered singular values of $\mM$ and $\mbs{\tilde{M}}$,
  Then, we have
  \begin{equation*}
    \sigma_i(\mbs{\tilde{M}}) \asympteq \sigma_i(\mM),\quad\forall i.
  \end{equation*}%
\end{proposition}
\begin{proof}  
  For $m\geq n$, we consider the left polar decomposition $\mM = \mU
  \mM_0$, where $\mU$ is a $m\times n$ matrix with orthonormal columns
  and $\mM_0$ a $n\times n$ positive definite matrix with
  $\sigma_i(\mM)=\sigma_i(\mM_0)$ for $i=1,\ldots,n$. Let
  $\mT\mU=\mV\mT_0$ be the left polar decomposition of $\mT\mU$. Then,
  we have $\sigma_i(\mbs{\tilde{M}})=\sigma_i(\mT_0\mM_0)$ for
  $i=1,\ldots,n$. 
  
  For $m<n$, we make a right polar decomposition $\mM = \mM_0
  \transc{\mU}$ , where $\mU$ is a $n\times m$ matrix with orthonormal
  columns and $\mM_0$ a $m\times m$ positive definite matrix with
  $\sigma_i(\mM)=\sigma_i(\mM_0)$ for $i=1,\ldots,n$. Then, we have
  $\sigma_i(\mbs{\tilde{M}})=\sigma_i(\mT_0\mM_0)$ for $i=1,\ldots,n$
  with $\mT_0\defeq\mT$.
  
  In both cases, the original problem is equivalent to showing that
  \begin{equation*}
    \sigma_i(\mT_0\mM_0) \asympteq \sigma_i(\mM_0),\quad\text{for}\ i=1,\ldots,q,
  \end{equation*}%
  with $\mT_0$ and $\mM_0$ now invertible. Let $\mA$ and $\mB$ in
  Lemma~\ref{lemma:lemma4} be $\mT_0$ and $\mM_0$, respectively. By
  applying \Eq{eq:sv:leq} and \Eq{eq:sv:geq} of appendix, we have
  \begin{equation*}
    \sigma_i(\mM_0)\sigma_m(\mT_0) \leq \sigma_i(\mbs{\tilde{M}}) \leq \sigma_i(\mM_0)\sigma_1(\mT_0), 
  \end{equation*}%
  from which we prove the proposition since
  $\sigma_1(\mT_0)\asympteq\sigma_m(\mT_0)\asympteq\SNR^0$ and
  $\sigma_i(\mM)=\sigma_i(\mM_0)$. 
\end{proof}
This proposition says that any invertible transformation with
bounded~(asymptotically in high SNR regime) eigenvalues does not
change the asymptotical p.d.f. of the singular values of a random
matrix.  According to this proposition, we know that the singular
values of $\PhiS^{1/2}\mH_1$ have the same asymptotical p.d.f. as the
ones of $\mH_1$, which leads to the main result of this work.
\begin{theorem}
\label{thm:3}
  For a $(\nT,\nS,\nR)$ double scattering MIMO channel
  \Eq{eq:channel-model} with $\mH$ defined in \Eq{eq:channel-matrix},
  the diversity-multiplexing tradeoff is a piecewise-linear function
  connecting the points $(k,d(k)),k=0,\ldots,M$ with $d(k)$ being
  defined in \Eq{eq:DMT-Prod-Rayleigh}, where $(M,N,L)$ is the ordered
  version of $(\nT,\nS,\nR)$ with $M\leq N\leq L$.
\end{theorem}
\begin{proof}
  This is a direct consequence of Theorem~\ref{thm:DMT-prod-Rayleigh},
  since the eigenvalues of
  $(\PhiS^{1/2}\mH_1)\transc{(\PhiS^{1/2}\mH_1)}$ have the same
  asymptotical p.d.f. as that of $\mH_1\transc{\mH}_1$.
\end{proof}
Note that all observations in Remark~\ref{remark:1} apply for the
general double scattering MIMO channel. In particular, the optimality
condition $L+1\geq M+N$ in observation~4 of Remark~\ref{remark:1} in
terms of $(\nT,\nS,\nR)$ is
\[2\max\left\{\nT,\nS,\nR\right\}+1\geq\nT+\nS+\nR,\] which is also the
condition under which the maximum channel diversity order $\nT \nS
\nR/\max\left\{\nT,\nS,\nR\right\}$ is achieved. Moreover, this
theorem implies that antenna or scatterer correlation does not,
indeed, have any impact on the DMT of a double scattering channel, as
long as the correlation matrices are non-singular. Finally, in the
singular correlation matrices case, it is straightfoward to show that
Theorem~\ref{thm:3} is still true, but with $(\nT,\nS,\nR)$ replaced
by $(\nT',\nS',\nR')$, the respective ranks of the correlation
matrices.

\section{Conclusion} 
\label{sec:conclusion}
We studied, in this paper, the DMT of a double scattering MIMO channel
and showed that, as long as the correlation matrices are non singular,
it is equal to the DMT of a Rayleigh MIMO product channel. This DMT is
always lower than the one of a single scattering ($\nT \times \nS$,
$\nS \times \nR$ or $\nT \times \nR$) MIMO channel and it is equal to
that one for certain values of the channel parameters.  This result is
not only interesting for itself, but it also helps to the calculation
of the DMT of MIMO Amplify-and-Forward \cite{SY_JCB_coop} cooperative
channels as the relayed link can be seen as a Rayleigh MIMO product
channel.

\appendix

\label{app:lemmas}
\begin{lemma}\label{lemma:lemma1}
  \begin{equation}
    \label{eq:detexp}
    \begin{split}
      &\Det\left[\exp\left(-\SNR^{-(\alpha_j-\beta_i)}\right)\right]_{i,j=1}^l  \\
      &\quad\asympteq
      \exp\left(-\sum_{i=1}^l\SNR^{-(\alpha_i-\beta_i)}\right)
      \SNR^{-\sum_{i<j}^l(\alpha_i-\beta_j)^+}.
    \end{split}
  \end{equation}%
\end{lemma}

\begin{proof}
  
  Let us define $D_l\defeq
  \Det\left[\exp\left(-\SNR^{-(\alpha_j-\beta_i)}\right)\right]_{i,j=1}^l$
  and we have
  \begin{equation*}
    \begin{split}
  D_l &= \Det\matrix{
     e^{-\SNR^{-(\alpha_1-\beta_1)}+\SNR^{-(\alpha_l-\beta_1)}} &\cdots &e^{-\SNR^{-(\alpha_{l-1}-\beta_1)}+\SNR^{-(\alpha_l-\beta_1)}} & 1\\ 
     \vdots & \ddots& \vdots & \vdots \\
     e^{-\SNR^{-(\alpha_1-\beta_l)}+\SNR^{-(\alpha_l-\beta_l)}} &\cdots &e^{-\SNR^{-(\alpha_{l-1}-\beta_l)}+\SNR^{-(\alpha_l-\beta_l)}} & 1 
     } e^{-\sum_i\SNR^{-(\alpha_l-\beta_i)}} \\
     &\asympteq \Det\matrix{
     e^{-\SNR^{-(\alpha_1-\beta_1)}}-e^{-\SNR^{-(\alpha_1-\beta_l)}} &\cdots& e^{-\SNR^{-(\alpha_{l-1}-\beta_1)}}-e^{-\SNR^{-(\alpha_{l-1}-\beta_l)}} & 0 \\
     \vdots & \ddots& \vdots & \vdots \\
     e^{-\SNR^{-(\alpha_1-\beta_{l-1})}}-e^{-\SNR^{-(\alpha_1-\beta_l)}} &\cdots& e^{-\SNR^{-(\alpha_{l-1}-\beta_{l-1})}}-e^{-\SNR^{-(\alpha_{l-1}-\beta_l)}} & 0 \\
     e^{-\SNR^{-(\alpha_1-\beta_{l})}} & \cdots & e^{-\SNR^{-(\alpha_{l-1}-\beta_{l})}} & 1
     } e^{-\SNR^{-(\alpha_l-\beta_l)}} \\
     &\asympteq \Det\matrix{
     e^{-\SNR^{-(\alpha_1-\beta_1)}}\left(1-e^{-\SNR^{-(\alpha_1-\beta_l)}}\right) &\cdots& e^{-\SNR^{-(\alpha_{l-1}-\beta_1)}}\left(1-e^{-\SNR^{-(\alpha_{l-1}-\beta_l)}}\right)\\
     \vdots & \ddots& \vdots\\
     e^{-\SNR^{-(\alpha_1-\beta_{l-1})}}\left(1-e^{-\SNR^{-(\alpha_1-\beta_l)}}\right) &\cdots& e^{-\SNR^{-(\alpha_{l-1}-\beta_{l-1})}}\left(1-e^{-\SNR^{-(\alpha_{l-1}-\beta_l)}}\right)
     } e^{-\SNR^{-(\alpha_l-\beta_l)}} \\
     &=  e^{-\SNR^{-(\alpha_l-\beta_l)}} \prod_{i=1}^{l-1}\left(1-e^{-\SNR^{-(\alpha_i-\beta_l)}}\right)D_{l-1}      
    \end{split}
  \end{equation*}
  where the equations are obtained by iterating the identity
  $\SNR^{-a}\pm\SNR^{-b} \asympteq \SNR^{-a}$ for $a<b$. Since
  $1-e^{-x}\approx x$ for $x$ close to $0^+$, we have
  $1-e^{-\SNR^{-(\alpha_i-\beta_l)}} \asympteq
  \SNR^{-(\alpha_i-\beta_l)}$ if $\alpha_i>\beta_l$ and
  $1-e^{-\SNR^{-(\alpha_i-\beta_l)}}\asympteq\SNR^0$ otherwise.  As
  shown in the recursive relation above, we must have
  $\alpha_i\geq\beta_i,\forall i$, in order that $D_l$ does not decay
  exponentially. Thus, we have $D_l \asympteq
  \SNR^{-\sum_{i<l}(\alpha_i-\beta_l)^+} D_{l-1}$, and in a recursive
  manner, we get \Eq{eq:detexp}.

\end{proof}

\begin{lemma}\label{lemma:lemma2}
  \begin{equation}
    \label{eq:lemma2}
    \begin{split}
      \Det\left(\mXi\right) &\asympteq \prod_{i=1}^{n+1}\SNR^{-(l+m-n-i)\alpha_i} \prod_{i=n+2}^{l}\SNR^{-(l+m+1-2i)\alpha_i} \nnb\\
      &\quad \cdot \prod_{i=1}^{n}\prod_{j=n+1}^{l}\SNR^{-(\alpha_i-\beta_j)^+}  \prod_{i<j}^n \SNR^{-(\alpha_i-\beta_j)^+} \nnb\\
      &\quad\cdot \exp\left(-\sum_{i=1}^n
        \SNR^{-(\alpha_i-\beta_i)}\right).
    \end{split}
  \end{equation}
\end{lemma}

\begin{proof}
  First, we have 
  \begin{equation}\label{eq:lemmas:tmp1}
    \Det{(\mXi)} = \prod_{i=1}^l \mu_i^{l-n-1} 
    \Det \matrix{
      \mu_1^{-(l-n-1)} &\cdots& 1 & e^{-\lambda_1/\mu_1}&\cdots&e^{-\lambda_n/\mu_1}\\
      \vdots & \ddots & \vdots& \vdots& \ddots & \vdots \\
      \mu_l^{-(l-n-1)} &\cdots& 1 & e^{-\lambda_1/\mu_l}&\cdots&e^{-\lambda_n/\mu_l}\\
    }.
  \end{equation}
  Then, let us denote the determinant in the right hand side of
  \Eq{eq:lemmas:tmp1} as $D$ and we rewrite it as
  \begin{align}
    D 
    &= \Det \matrix{
      d_{1,l}^{(l-n-1)} &\cdots& 0 & e^{-\lambda_1/\mu_1}-e^{-\lambda_1/\mu_l}&\cdots&e^{-\lambda_n/\mu_1}-e^{-\lambda_n/\mu_l}\\
      \vdots & \ddots & \vdots& \vdots& \ddots & \vdots \\
      d_{l-1,l}^{(l-n-1)} &\cdots& 0 & e^{-\lambda_1/\mu_{l-1}}-e^{-\lambda_1/\mu_{l}}&\cdots&e^{-\lambda_n/\mu_{l-1}}-e^{-\lambda_n/\mu_{l}}\\
      \mu_l^{-(l-n-1)} &\cdots& 1 & e^{-\lambda_1/\mu_l}&\cdots&e^{-\lambda_n/\mu_l}\\
    } \label{eq:lemmas:tmp3} \\
    &\asympteq \Det \matrix{
      d_{1,l}^{(l-n-1)} &\cdots& d_{1,l}^{(1)} & e^{-\lambda_1/\mu_1}&\cdots&e^{-\lambda_n/\mu_1}\\
      \vdots & \ddots & \vdots& \vdots& \ddots & \vdots \\
      d_{l-1,l}^{(l-n-1)} &\cdots& d_{l-1,l}^{(1)} & e^{-\lambda_1/\mu_{l-1}}&\cdots&e^{-\lambda_n/\mu_{l-1}}\\
    } \prod_{i=1}^n \left(1-e^{-\lambda_i/\mu_l}\right) \label{eq:lemmas:tmp2}
  \end{align}%
  where $d_{i,j}^{(k)}\defeq \mu_i^{-k} - \mu_j^{-k}$ and the product
  term in \Eq{eq:lemmas:tmp2} is obtained since
  $1-e^{-(\lambda_i/\mu_l-\lambda_i/\mu_j)}\asympteq
  1-e^{-\lambda_i/\mu_l}$ for all $j<l$. Let us denote the determinant
  in \Eq{eq:lemmas:tmp2} as $D_l$. Then, by multiplying the first
  column in $D_l$ with $\mu_l^{l-n-1}$ and noting that $\mu_l^{l-n-1}
  d_{i,l}^{(l-n-1)}=1-\left(\D\frac{\mu_l}{\mu_i}\right)^{l-n-1}\approx
  1$, the first column of $D_l$ becomes all $1$. Now, by eliminating
  the first $l-2$ ``$1$''s of the first column by substracting all
  rows by the last row as in \Eq{eq:lemmas:tmp3} and
  \Eq{eq:lemmas:tmp2}, we have $\mu_l^{l-n-1} D_l\asympteq
  \prod_{i=1}^n \left(1-e^{-\lambda_i/\mu_l}\right) D_{l-1}$. By
  continuing reducing the dimension, we get
  \begin{equation*}
    \begin{split}
      \Det(\mXi) &\asympteq  \Det\left[e^{-\lambda_j/\mu_i}\right]_{i,j=1}^n \prod_{i=1}^{n+1}\mu_i^{l-n-1}\prod_{i=n+2}^l \mu_i^{l-i}\\
      &\quad \cdot\prod_{i=1}^n\prod_{j=n+1}^l
      \left(1-e^{-\lambda_i/\mu_j}\right)
    \end{split}
  \end{equation*}
  from which we prove the lemma, by applying \Eq{eq:detexp}.  
\end{proof}

\begin{lemma}\label{lemma:lemma3}
  Let $\mW$ be a central complex Wishart matrix
  $\mW\sim\Wcal_m(n,\mSigma)$ with $n<m$, where the ordered
  eigenvalues of $\mSigma$ are
  $\mu_1>\ldots>\mu_{l}>\mu_{l+1}=\ldots=\mu_m=0$ with $l\geq n$. The
  joint p.d.f. of the ordered positive eigenvalues
  $\lambda_1>\ldots>\lambda_n$ of $\mW$ equals
  \begin{equation}
    \label{eq:Wishart:n<m2}
    G_{m,n} \Det(\mXi_l) \prod_{i<j}^l \frac{1}{(\mu_i-\mu_j)} \prod_{i<j}^n (\lambda_i-\lambda_j) 
  \end{equation}%
  with 
  \begin{equation}
    \label{eq:def-Xit}
    \mXi_l \defeq \matrix{1 & \mu_1 & \cdots & \mu_1^{l-n-1} & \mu_1^{l-n-1}e^{-\frac{\lambda_1}{\mu_1}} & \cdots & \mu_1^{l-n-1}e^{-\frac{\lambda_n}{\mu_1}} \\
      \vdots & \vdots & \ddots &\vdots& \vdots& \ddots & \vdots \\
        1 & \mu_l & \cdots & \mu_l^{l-n-1} & \mu_l^{l-n-1}e^{-\frac{\lambda_1}{\mu_l}} & \cdots & \mu_l^{l-n-1}e^{-\frac{\lambda_n}{\mu_l}}
}. 
  \end{equation}%
\end{lemma}
\vspace{0.2cm}
\begin{proof}
  We can prove it by successively applying the l'Hospital rule with
  $\mu_m,\ldots,\mu_{l+1}\to 0$ on the expression in
  \Eq{eq:Wishart:n<m}. Let us prove by induction that 
  \begin{equation}\label{eq:lemmas:tmp4}
    \lim_{\mu_{l+1},\ldots,\mu_m \to 0}\frac{\Det(\mXi)}{\prod_{i<j}^m (\mu_i-\mu_j)} = \frac{\Det(\mXi_l)}{\prod_{i<j}^l (\mu_i-\mu_j)}. 
  \end{equation}%
  For $l=m-1$, \Eq{eq:lemmas:tmp4} is obviously true. Then, assuming
  that \Eq{eq:lemmas:tmp4} holds for given $l$, then, as long as
  $l-1\geq n$, we have
  \begin{eqnarray}
    \lim_{\mu_{l},\ldots,\mu_m \to 0}\frac{\Det(\mXi)}{\prod_{i<j}^m (\mu_i-\mu_j)} 
    &=& \lim_{\mu_{l} \to 0}\frac{\Det(\mXi_l)}{\prod_{i<j}^l (\mu_i-\mu_j)} \nnb\\
    &=& \frac{\Det(\mXi_{l-1})}{\prod_{i<j}^{l-1} (\mu_i-\mu_j)} \label{eq:lemmas:tmp5} 
  \end{eqnarray}%
  where \Eq{eq:lemmas:tmp5} is deduced from \Eq{eq:lemmas:tmp4}. 
\end{proof}

\begin{lemma}\label{lemma:lemma4}
  Let $\mA$ and $\mB$ be two $m\times m$ non-singular matrices. For any
  $n\times n$ matrix $\mM$, let $\sigma_i(\mM)$ be the $i$th largest
  singular value of $\mM$ and $\eta_i{(\mM)}$ be the $i$th smallest
  one~(\ie, $\sigma_i(\mM)=\eta_{n+1-i}(\mM)$). Then, we have
  \begin{eqnarray}
    \sigma_{i+j-1}(\mA\mB) &\leq& \sigma_i(\mA)\,\sigma_j(\mB) \label{eq:sv:leq}\\
    \eta_{i+j-1}(\mA\mB)   &\geq& \eta_i(\mA)\,\eta_j(\mB) \label{eq:sv:geq}
  \end{eqnarray}%
  for $1\leq \{i,j\} \leq m$ and $i+j\leq m+1$.
\end{lemma}
\begin{proof}
  Let $\mA\mB = \mU \mQ$ be the left polar decomposition of $\mA\mB$
  with $\mU$ unitary and $\mQ$ positive definite. Then, we have
  $\mQ=\transc{\mU}\mA\mB$ and $\sigma_i(\mQ)=\sigma_i(\mA\mB),\forall
  i$. The quadratic form $\transc{\mx}\,\mQ\mx$ can be bounded as
  \begin{equation}
    \label{eq:lemmas:tmp7}
    \begin{split}
      \Abssqr{\transc{\mx}\mQ\mx}
      &= \Abssqr{\transc{\left(\transc{\mA}\mU\mx\right)}\left(\mB\mx\right)} \\
      &\leq \Norm{\transc{\mA}\,\mU\mx}^2\Norm{\mB\mx}^2 \\
      &= \left(\transc{\mx}_1\mQ_{\mA}\mx_1\right)
      \left(\transc{\mx}\,\mQ_{\mB}\mx\right)
    \end{split}
  \end{equation}%
  where $\mx_1 \defeq \mU \mx$, $\mQ_{\mA} \defeq \mA\transc{\mA}$ and
  $\mQ_{\mB} \defeq \mB\transc{\mB}$. The eigenvalue decomposition of
  $\mQ_{\mA}$ and $\mQ_{\mB}$ gives
  \begin{align*}
    \mQ_{\mA} &= \sum_{i=1}^m \sigma_i^2(\mA) \mz_i \transc{\mz}_i\quad\text{and}\\
    \mQ_{\mB} &= \sum_{i=1}^m \sigma_i^2(\mB) \my_i \transc{\my}_i
  \end{align*}%
  where $\mz_i$ and $\my_i$ are eigenvectors of $\mQ_{\mA}$ and
  $\mQ_{\mB}$, respectively. Now, taking $\mx_k = \transc{\mU} \mz_k$
  for $k=1,\ldots,i-1$ and $\mx_k = \my_{k-i+1}$ for
  $k=i,\ldots,i+j-2$, we have, $\forall \mx \perp \mx_k$ for
  $k=1,\ldots,i+j-2$,
  \begin{align}
    \transc{\left(\mU\mx\right)}\,\mQ_{\mA}\left(\mU\mx\right) &\leq
    \sigma_i^2(\mA) \Norm{\mx}^2 \label{eq:lemmas:tmp8}\\
    \transc{\mx}\,\mQ_{\mB}\mx &\leq \sigma_j^2(\mB) \Norm{\mx}^2. \label{eq:lemmas:tmp9}
  \end{align}%
  From \Eq{eq:lemmas:tmp7}, \Eq{eq:lemmas:tmp8} and
  \Eq{eq:lemmas:tmp9} and the Courant-Fischer theorem~\cite{Horn}, we
  get
  \begin{equation*}
    \sigma_i^2(\mA\mB) \leq \max_{\mx\perp\,\mx_1,\ldots,\mx_{i+j-2}} \frac{\Abssqr{\transc{\mx}\,\mQ\mx}}{\Norm{\mx}^4} \leq \sigma_i^2(\mA)\sigma_j^2(\mB),
  \end{equation*}%
  from which we have \Eq{eq:sv:leq}. 
  
  Note that for any invertible matrix $\mM$, we have
  $\eta_i(\mM)=\sigma_i^{-1}(\inv{\mM})$. By applying this equality
  and using the inequality \Eq{eq:sv:leq}, it is straightfoward to get
  \Eq{eq:sv:geq} after some simple manipulations.

\end{proof}

\end{document}